# Fluctuation-Dissipation Inequality for Quadratic Open Systems

*Abstract*

For open systems derived from quadratic total Hamiltonians, we derive a dynamic fluctuation-dissipation (FD) inequality valid for any total initial state and without regard to the sign of the dissipation. With the added constraint that this state be factorized with the reservoir in thermal equilibrium, an uncertainty relation arises naturally from the FD inequality that can be stronger than the usual uncertainty principle in the form $\langle q^2 \rangle \langle p^2 \rangle \geq \hbar^2/4$. We discuss some of the properties of the uncertainty relation relevant to decoherence.

Allan Tameshtit; University of Toronto, St. George Campus, Toronto, Ontario, Canada;
Telephone number: 416-460-2270; email: atameshtit@utoronto.ca

1. *Introduction*

Quadratic Hamiltonian models play a special role in the study of quantum open systems because they are among the few that can be solved analytically. The class of quantum open systems that will be studied here derives from a widely-used microscopic model analyzed extensively in the 1960s by Ullersma [1]. In this model, the total system comprises an oscillator (system of interest) bilinearly coupled to a large number of other oscillators (reservoir). The literature pertaining to Ullersma's model is vast. Refs. [2,3], dealing with the uncertainty principle, are especially related to the work presented below.

In this letter, we derive a dynamic fluctuation-dissipation (FD) relation that, unlike the standard FD theorem, involves an inequality in the time domain. The derivation of the FD inequality proceeds without reference to any particular initial state of the total system and without regard to the sign of the dissipation. In some recent work that emphasized stationary correlations, a FD inequality was also obtained but in the frequency domain [4].

As an application of the FD inequality derived herein, we constrain ourselves to a factorized total initial state with the reservoir in thermal equilibrium,

$$\rho_T(0) = \rho \otimes \rho_{res,eq}. \tag{1}$$

This state naturally leads us to define below a function of second moments, $D(t)$, which on account of the FD inequality, obeys an uncertainty relation that can be stronger than one form of the conventional uncertainty principle, $\langle q^2 \rangle \langle p^2 \rangle - \hbar^2/4 \geq 0$. This uncertainty relation, $D(t) \geq 0$, is related to some earlier work [2,3,5] that exploited the latter and other forms of the uncertainty principle to derive various inequalities. We will also examine the connection between the uncertainty relation $D(t) \geq 0$ and some previous results [6] involving positivity properties of a class of master equations. We finally discuss some of the properties of the uncertainty relation relevant to decoherence.

2. *Fluctuation-Dissipation Inequality from Ullersma's Model*

We consider Ullersma's model, where the total Hamiltonian is given by
$$H_T = \sum_{\nu=0}^{N} \frac{1}{2}\left(P_\nu^2 + \omega_\nu^2 Q_\nu^2\right) + \sum_{n=1}^{N} \varepsilon_n Q_0 Q_n, \qquad (2)$$
where $(Q_0, P_0)$ and $(Q_1,...,Q_N, P_1,...,P_N)$ are the mass-weighted canonical coordinates of the system of interest and reservoir, respectively: $P_\nu = p_\nu / m_\nu^{1/2}$ and $Q_\nu = m_\nu^{1/2} q_\nu$, $\nu = 0,1,2,...$ We assume that $\sum_{n=1}^{N} \varepsilon_n^2 / \omega_n^2$ is bounded and, to guarantee that $H_T$ is positive-definite, that moreover [1]
$$\sum_{n=1}^{N} \frac{\varepsilon_n^2}{\omega_n^2} \leq \omega_0^2.$$
Using a normal mode analysis, Ullersma solved the equations of motion governed by $H_T$ for the system of interest, obtaining:
$$Q_0(t) = \sum_{\nu=0}^{N} \left[\dot{A}_{0\nu}(t) Q_\nu(0) + A_{0\nu}(t) P_\nu(0)\right] \qquad (3)$$
$$P_0(t) = \sum_{\nu=0}^{N} \left[\ddot{A}_{0\nu}(t) Q_\nu(0) + \dot{A}_{0\nu}(t) P_\nu(0)\right] \qquad (4)$$
where, letting $A(t) \equiv A_{00}(t)$ and $n = 1,2,...$,
$$A_{0n}(t) = -\frac{\varepsilon_n}{\omega_n} \int_0^t A(t') \sin \omega_n(t-t') dt', \qquad (5)$$
and
$$A(t) = \frac{1}{2\pi i} \oint dz \frac{\sin(zt)}{g(z)}, \qquad (6)$$
with
$$g(z) = z^2 - \omega_0^2 - \sum_{n=1}^{N} \frac{\varepsilon_n^2}{z^2 - \omega_n^2}$$ and where the closed contour encircles the positive real axis [1,7].

Defining $R(t) \equiv \sqrt{\dot{A}^2(t) - A(t)\ddot{A}(t)}$, Eqs. (3), (4) and the commutation relation $[Q_0(t), P_0(t)] = \hbar i$ imply
$$1 - R^2(t) = \sum_{n=1}^{} \left(\dot{A}_{0n}^2(t) - A_{0n}(t)\ddot{A}_{0n}(t)\right)$$
$$= \sum_{n=1} \varepsilon_n^2 \left\{ \left[\int_0^t A(t')\cos[\omega_n(t-t')]dt'\right]^2 + \left[\int_0^t A(t')\sin[\omega_n(t-t')]dt'\right]^2 - \frac{A(t)}{\omega_n}\int_0^t A(t')\sin[\omega_n(t-t')]dt' \right\}$$
$$(7)$$
where this last expression is obtained by using Eq. (5).

We now introduce the quantity

$$F_E(\omega_n,t) = -\frac{\varepsilon_n}{\omega_n}\sqrt{E(\omega_n,x)}\int_0^t A(t')\exp[i\omega_n(t-t')]dt' \tag{8}$$

where $E(\omega_n,x)$ is a function, having units of energy, satisfying

$$E(\omega_n,0) = \hbar\omega_n/2, \tag{9}$$

$$E(\omega_n,x) \geq 0 \text{ and } \partial E(\omega_n,x)/\partial x \geq 0 \text{ [8]}. \tag{10}$$

The main motivation for defining $F_E$ in this way is that if we look ahead, a particular member of this class of functions $F_E$, with $x$ and $E(\omega_n,x)$ equal to the temperature and $\frac{\hbar\omega_n}{2}\coth\frac{\hbar\omega_n}{2kT}$ respectively, arises when computing various expectation values as the total system evolves from the initial state given by (1). At this point, however, $x$ is an arbitrary non-negative parameter, and no reference is made to any particular initial total state.

Considering $(F_E(\omega_1), F_E(\omega_2),..., F_E(\omega_N)) \in \mathbb{C}^N$, the Cauchy-Schwarz inequality yields

$$\sum_{m=1}^{N}|F_E(\omega_m)|^2 \sum_{n=1}^{N}\left|\frac{\partial F_E(\omega_n)}{\partial t}\right|^2 \geq \left|\sum_{n=1}^{N} F_E^*(\omega_n)\frac{\partial F_E(\omega_n)}{\partial t}\right|^2 \tag{11}$$

$$= \frac{1}{4}\left[\frac{\partial}{\partial t}\sum_{m=1}^{N}|F_E(\omega_m)|^2\right]^2 + \left[\sum_{n=1}^{N}\text{Im}\left(F_E^*(\omega_n)\frac{\partial F_E(\omega_n)}{\partial t}\right)\right]^2 \tag{12}$$

where this last expression is obtained by using the identity $\text{Re}\left(F^*\frac{\partial F}{\partial t}\right) = \frac{1}{2}\frac{\partial}{\partial t}\left(|F|^2\right)$.

The sum in the last term of Eq. (12) may be written as

$$\sum_{n=1}^{N}\text{Im}\left(F_E^*(\omega_n)\frac{\partial F_E(\omega_n)}{\partial t}\right) =$$

$$\hbar\sum_{n=1}^{N}\varepsilon_n^2\left(\frac{E(\omega_n,x)}{\hbar\omega_n}-\frac{1}{2}\right)\left\{\left[\int_0^t A(t')\cos[\omega_n(t-t')]dt'\right]^2 + \left[\int_0^t A(t')\sin[\omega_n(t-t')]dt'\right]^2 - \frac{A(t)}{\omega_n}\int_0^t A(t')\sin[\omega_n(t-t')]dt'\right\}$$

$$+\frac{\hbar}{2}(1-R^2(t)), \tag{13}$$

where we have used Eq. (7). Substituting this last expression into relation (12), we obtain

$$\sum_{m=1}^{N}|F_E(\omega_m)|^2 \sum_{n=1}^{N}\left|\frac{\partial F_E(\omega_n)}{\partial t}\right|^2 - \frac{1}{4}\left[\frac{\partial}{\partial t}\sum_{n=1}^{N}|F_E(\omega_n)|^2\right]^2$$

$$\geq \left\{\hbar\sum_{n=1}^{N}\varepsilon_n^2\left(\frac{E(\omega_n,x)}{\hbar\omega_n}-\frac{1}{2}\right)\left\{\left[\int_0^t A(t')\cos[\omega_n(t-t')]dt'\right]^2 + \left[\int_0^t A(t')\sin[\omega_n(t-t')]dt'\right]^2 - \frac{A(t)}{\omega_n}\int_0^t A(t')\sin[\omega_n(t-t')]dt'\right\} \right.$$

$$\left. +\frac{\hbar}{2}(1-R^2(t)) \right\}^2$$

(14)

Therefore, on account of assumption (9), the inequality

$$\sum_{m=1}^{N}|F_E(\omega_m)|^2 \sum_{n=1}^{N}\left|\frac{\partial F_E(\omega_n)}{\partial t}\right|^2 - \frac{1}{4}\left[\frac{\partial}{\partial t}\sum_{n=1}^{N}|F_E(\omega_n)|^2\right]^2 - \frac{\hbar^2}{4}(1-R^2(t))^2 \geq 0 \qquad (15)$$

is seen to be valid at $x=0$, and we will now show that it is valid for all $x \geq 0$ under assumptions (10).

After differentiating with respect to $x$ and integrating by parts with respect to time, the resultant time-dependent factors can be placed in a manifestly non-negative form [9] to yield:

$$\frac{\partial}{\partial x}\left\{\sum_{m=1}^{N}|F_E(\omega_m)|^2 \sum_{n=1}^{N}\left|\frac{\partial F_E(\omega_n)}{\partial t}\right|^2 - \frac{1}{4}\left[\frac{\partial}{\partial t}\sum_{n=1}^{N}|F_E(\omega_n)|^2\right]^2 - \frac{\hbar^2}{4}(1-R^2(t))^2\right\} =$$

$$\sum_{m=1}^{N}\sum_{n=1}^{N}\left(\frac{\varepsilon_m \varepsilon_n}{\omega_m \omega_n}\right)^2 E(\omega_m, x)\frac{\partial E(\omega_n, x)}{\partial x} \times$$

$$\left\{\begin{array}{l}[C_0(\omega_n)C_1(\omega_m) - C_0(\omega_m)C_1(\omega_n)]^2 + [S_1(\omega_n)C_0(\omega_m) - S_0(\omega_n)C_1(\omega_m)]^2 \\ + [S_1(\omega_m)C_0(\omega_n) - S_0(\omega_m)C_1(\omega_n)]^2 + [S_1(\omega_m)S_0(\omega_n) - S_1(\omega_n)S_0(\omega_m)]^2\end{array}\right\} \geq 0, \qquad (16)$$

where we define the time dependent functions $C_0(\omega) = \int_0^t A(t')\cos\omega(t-t')dt'$ and $C_1(\omega) = \int_0^t \dot{A}(t')\cos\omega(t-t')dt'$, with similar expressions for $S_0(\omega)$ and $S_1(\omega)$ but with the cosine function replaced by sine, and we note that the last inequality in expression (16) is true on account of assumptions (10). Thus, the left-hand side of Eq. (15) is a non-decreasing function of $x$, and since we have seen that inequality (15) is valid at $x=0$, we conclude that it is valid for all $x \geq 0$.

We will see that under certain circumstances, it is natural to associate $x$ with the temperature. The first two terms on the left-hand side of relation (15) depend on $x$ and are related to the fluctuations caused by the reservoir. The last term, independent of $x$, is related to dissipation. For this reason we consider expression (15) to be a type of dynamic fluctuation-dissipation relation.

3. *Application: Derivation of an Uncertainty Relation*

In the derivation of inequality (15), we did not reference any particular initial state for the total system. To obtain an uncertainty relation involving the second moments and to make contact with the literature, in this section we consider the total initial state given by expression (1). In such case, it is natural to set $x = T$ and to choose the energy function $E(\omega_n, T) = \frac{\hbar\omega_n}{2}\coth\frac{\hbar\omega_n}{2kT}$, the well-known mean energy of a reservoir oscillator in thermal equilibrium at temperature $T$. Inequality (15) becomes

$$XY - \frac{1}{4}\dot{X}^2 - \frac{\hbar^2}{4}(1-R^2)^2 \geq 0 \qquad (17)$$

where

$$X(t) = \frac{\hbar}{2}\sum_{n=1}^{N}\frac{\varepsilon_n^2}{\omega_n}\coth\left(\frac{\hbar\omega_n}{2kT}\right)\left|\int_0^t e^{i\omega_n t'}A(t')dt'\right|^2 \qquad (18)$$

and

$$Y(t) = \frac{\hbar}{2} \sum_{n=1}^{N} \frac{\varepsilon_n^2}{\omega_n} \coth\left(\frac{\hbar \omega_n}{2kT}\right) \left| \int_0^t e^{i\omega_n t'} A(t') dt' \right|^2 \qquad (19)$$

In Ref. [2], a related inequality was provided after considering a factorized initial state with the reservoir in thermal equilibrium and applying the uncertainty principle. This inequality, expression (4.18) of Ref. [2], and statements therein imply $XY - \dot{X}^2/4 \geq \hbar^2(1-R^4)/4$, where we have used our notation and gone to the continuous frequency case. This result is similar but not identical to expression (17) [11]. We may consider expression (17) to be an extension to non-autonomous systems of an inequality derived previously for master equations with time-independent coefficients (inequality (11) in Ref. [5]). From the work in Ref. [13], it follows that relation (17) together with $X(t) \geq 0$ and $Y(t) \geq 0$ are sufficient to ensure that $\rho(t)$ is positive.

To formulate from inequality (17) an uncertainty relation involving second moments, we make use of the following expressions obtained in Ref. [7]:

$$\langle q_0^2 \rangle(t) = \frac{X(t)}{m_0} + \dot{A}^2(t)\langle q_0^2 \rangle(0) + \frac{A(t)\dot{A}(t)}{m_0}\langle q_0 p_0 + p_0 q_0 \rangle(0) + \frac{A^2(t)}{m_0^2}\langle p_0^2 \rangle(0) \qquad (20)$$

$$\langle p_0^2 \rangle(t) = m_0 Y(t) + m_0^2 \ddot{A}^2(t)\langle q_0^2 \rangle(0) + m_0 \dot{A}(t)\ddot{A}(t)\langle q_0 p_0 + p_0 q_0 \rangle(0) + \dot{A}^2(t)\langle p_0^2 \rangle(0) \qquad (21)$$

$$\langle q_0 p_0 + p_0 q_0 \rangle(t) = \dot{X}(t) + 2m_0 \dot{A}(t)\ddot{A}(t)\langle q_0^2 \rangle(0) + \left(\dot{A}^2(t) + A(t)\ddot{A}(t)\right)\langle q_0 p_0 + p_0 q_0 \rangle(0)$$
$$+ \frac{2}{m_0} A(t)\dot{A}(t)\langle p_0^2 \rangle(0) \qquad (22)$$

Supposing $\dot{A}^2(t) - A(t)\ddot{A}(t) \equiv R^2(t) > 0$, we can re-write the second half of these expressions as follows:

$$\dot{A}^2(t)\langle q_0^2 \rangle(0) + \frac{A(t)\dot{A}(t)}{m_0}\langle q_0 p_0 + p_0 q_0 \rangle(0) + \frac{A^2(t)}{m_0^2}\langle p_0^2 \rangle(0) = R^2(t)Tr(q_0^2 \tilde{U}(t)\rho \tilde{U}^\dagger(t)) \qquad (23)$$

$$m_0^2 \ddot{A}^2(t)\langle q_0^2 \rangle(0) + m_0 \dot{A}(t)\ddot{A}(t)\langle q_0 p_0 + p_0 q_0 \rangle(0) + \dot{A}^2(t)\langle p_0^2 \rangle(0) = R^2(t)Tr(p_0^2 \tilde{U}(t)\rho \tilde{U}^\dagger(t)) \qquad (24)$$

$$2m_0 \dot{A}(t)\ddot{A}(t)\langle q_0^2 \rangle(0) + \left(\dot{A}^2(t) + A(t)\ddot{A}(t)\right)\langle q_0 p_0 + p_0 q_0 \rangle(0) + \frac{2}{m_0} A(t)\dot{A}(t)\langle p_0^2 \rangle(0) =$$
$$R^2(t)Tr((q_0 p_0 + p_0 q_0)\tilde{U}(t)\rho \tilde{U}^\dagger(t)) \qquad (25)$$

where the unitary propagator $\tilde{U}(t)$ is defined via the following action (see [6] and also cf. Eqs. (3.18) in Ref. [2]),

$$\tilde{U}^{\dagger}(t)q_0\tilde{U}(t)=\frac{1}{R(t)}\left(\dot{A}(t)q_0+\frac{A(t)}{m_0}p_0\right) \tag{26}$$

$$\tilde{U}^{\dagger}(t)p_0\tilde{U}(t)=\frac{1}{R(t)}\left(m_0\ddot{A}(t)q_0+\dot{A}(t)p_0\right) \tag{27}$$

Using inequality (17) and Eqs. (20)-(25), we obtain the uncertainty relation

$$(\delta q)^2(\delta p)^2-\frac{1}{4}C_\delta^2\geq\frac{\hbar^2}{4}(1-R^2)^2 \tag{28}$$

where

$$\delta q(t)=[\langle q_0^2\rangle(t)-R^2(t)Tr(q_0^2\tilde{U}(t)\rho\tilde{U}^{\dagger}(t))]^{1/2}, \tag{29}$$

$$\delta p(t)=[\langle p_0^2\rangle(t)-R^2(t)Tr(p_0^2\tilde{U}(t)\rho\tilde{U}^{\dagger}(t))]^{1/2} \text{ and} \tag{30}$$

$$C_\delta(t)=\langle q_0p_0+p_0q_0\rangle(t)-R^2(t)Tr((q_0p_0+p_0q_0)\tilde{U}(t)\rho\tilde{U}^{\dagger}(t)). \tag{31}$$

The two quantities in square brackets in Eqs. (29) and (30), equal to $X/m_0$ and $m_0Y$ respectively, are non-negative. Consequently, $\delta q$ and $\delta p$ are real and non-negative.

Inequality (28) can be compared with the conventional Robertson-Schrodinger uncertainty principle

$$(\Delta q)^2(\Delta p)^2-\frac{1}{4}C_\Delta^2\geq\frac{\hbar^2}{4} \tag{32}$$

where $C_\Delta=\langle qp+pq\rangle-2\langle q\rangle\langle p\rangle$.

4. *Relation to Previous Work*

Inequality (17) was derived in a different manner in [6] under various assumptions that include the dissipation being positive for $t>0$ and the reduced density operator being a positive operator (a condition which may be in doubt if approximations are made to derive the reduced density operator). In contrast, neither of these two conditions was assumed in the derivation presented above, the main assumption there instead being that the commutation relation $[Q_0(t),P_0(t)]=\hbar i$ holds for all time. In this section, we examine why these two conditions were necessary in the previous work, and to this end, we first briefly review the main results therein, which centered on quantum open systems described by the following type of master equation [14]:

$$\frac{d}{dt}\rho(t)=\frac{1}{\hbar i}[H_s(t),\rho(t)]-k_1(t)\{q,\rho(t),q\}-k_2(t)\{p,\rho(t),p\}+k_3(t)\{p,\rho(t),q\}+k_4(t)\{q,\rho(t),p\}$$
$$\tag{33}$$

where $H_s(t) \equiv b_{11}(t)q^2 + b_{12}(t)(qp+pq) + b_{22}(t)p^2$, with $b_{11}, b_{12}, b_{22}, k_1$ and $k_2$ being real continuous functions of time, $k_3$ being a complex continuous function of time, $k_4 = k_3^*$ and $\{A, \rho, B\} \equiv BA^\dagger \rho + \rho BA^\dagger - 2A^\dagger \rho B$.

Eq. (33) has the following operator solution:

$$\rho(t) = \exp\left[-\frac{w_4}{e^{w_4}-1}\left(w_1\{q,\cdot,q\} + w_2\{p,\cdot,p\} - \left(w_3 - i\frac{e^{w_4}-1}{4\hbar}\right)\{p,\cdot,q\} - \left(w_3 + i\frac{e^{w_4}-1}{4\hbar}\right)\{q,\cdot,p\}\right)\right]\mathcal{U}_{rev}\rho(0) \quad (34)$$

where the operator $\mathcal{U}_{rev}$, giving rise to reversible evolution, satisfies $\frac{d\mathcal{U}_{rev}}{dt} = \mathcal{L}_s \mathcal{U}_{rev}$ with $\mathcal{L}_s = \frac{1}{\hbar i}[H_s(t),\cdot]$ and initial condition $\mathcal{U}_{rev}(0) = 1$, and where the coefficients in the preceding exponent are solutions to the following inhomogeneous system of differential equations:

$$\frac{d}{dt}\begin{pmatrix} w_1 \\ w_2 \\ w_3 \end{pmatrix} = 2\begin{pmatrix} -2b_{12} & 0 & -2b_{11} \\ 0 & 2b_{12} & 2b_{22} \\ b_{22} & -b_{11} & 0 \end{pmatrix}\begin{pmatrix} w_1 \\ w_2 \\ w_3 \end{pmatrix} + e^{w_4}\begin{pmatrix} k_1 \\ k_2 \\ (k_3+k_4)/2 \end{pmatrix} \quad (35)$$

with

$$w_4 = 2\hbar i \int_0^t (k_3 - k_4)dt', \quad (36)$$

and initial conditions $w_1(0) = w_2(0) = w_3(0) = 0$ [6].

In addition to the foregoing conditions, suppose that for $t \geq 0$,

$w_1(t) \geq 0$, $w_2(t) \geq 0$ and $\rho(0)$ is an allowable initial density operator. (37)

Suppose further that for $t > 0$,

$w_1(t) > 0, w_4(t) > 0$ and $w_1(t)w_2(t) - w_3^2(t) \geq 0$. (38)

According to a theorem in Refs. [6,12], it follows that if $\rho(t)$ is positive for all $t \geq 0$, then

$$w_1(t)w_2(t) - w_3^2(t) \geq \left(\frac{e^{w_4(t)}-1}{4\hbar}\right)^2 \quad (39)$$

for all $t \geq 0$, which corresponds to inequality (17). On a related note, it is known that $-r_1\{q,\cdot,q\} + r_3\{q,\cdot,p\} + r_3^*\{p,\cdot,q\} - r_2\{p,\cdot,p\}$ can be written in Lindblad form, $-\sum_n \{a_n q + b_n p, \cdot, a_n q + b_n p\}$, with complex coefficients $a_n$ and $b_n$, if $r_1 \geq 0$, $r_2 \geq 0$ and $r_1 r_2 \geq |r_3|^2$ [15]. Consequently, relation (39), together with $w_1(t) \geq 0$ and $w_2(t) \geq 0$, are conditions that ensure the exponent in Eq. (34) can be written in Lindblad form. Unlike Eq. (33), however, the master equations of interest in Lindblad's article [16] are autonomous.

The second moments are given by [6,12]

$$\langle q^2 \rangle(t) \equiv Tr q^2 \rho(t)$$
$$= e^{-w_4(t)} [Tr q^2 \mathcal{U}_{rev}(t)\rho(0) + 2\hbar^2 w_2(t)], \tag{40}$$

$$\langle p^2 \rangle(t) = e^{-w_4(t)} [Tr p^2 \mathcal{U}_{rev}(t)\rho(0) + 2\hbar^2 w_1(t)] \tag{41}$$

and

$$\langle qp + pq \rangle(t) = e^{-w_4(t)} [Tr(qp + pq)\mathcal{U}_{rev}(t)\rho(0) + 4\hbar^2 w_3(t)]. \tag{42}$$

Combining expressions (39)-(42) results in

$$[\langle q^2 \rangle(t) - e^{-w_4(t)} Tr q^2 \mathcal{U}_{rev}(t)\rho(0)] [\langle p^2 \rangle(t) - e^{-w_4(t)} Tr p^2 \mathcal{U}_{rev}(t)\rho(0)]$$
$$- \frac{1}{4}[\langle qp + pq \rangle(t) - e^{-w_4(t)} Tr(qp + pq)\mathcal{U}_{rev}(t)\rho(0)]^2 \geq \frac{\hbar^2}{4}(1 - e^{-w_4(t)})^2, \tag{43}$$

which corresponds to the uncertainty relation (28).

For the factorized total state (1), it is instructive to examine the conditions for Ullersma's model that correspond to assumptions (37) and (38). Assuming $R^2 > 0$, the evolution of $\rho(t)$ governed by $H_T$ is given by [13]

$$\rho(t) = \exp\left[\frac{\ln R^2}{2\hbar^2(1-R^2)}\begin{pmatrix} m_0 Y\{q,\cdot,q\} + \frac{X}{m_0}\{p,\cdot,p\} \\ -\frac{1}{2}(\dot{X} - i\hbar(1-R^2))\{p,\cdot,q\} - \frac{1}{2}(\dot{X} + i\hbar(1-R^2))\{q,\cdot,p\} \end{pmatrix}\right] \tilde{U}(t)\rho(0)\tilde{U}^\dagger(t), \tag{44}$$

which result is the operator version of the Wigner function solution obtained in [7]. In the interaction picture where $\rho'(t) \equiv \tilde{U}^\dagger(t)\rho(t)\tilde{U}(t)$ and assuming $XY - \dot{X}^2/4 > 0$ for $t > 0$, the solution may also be written in terms of generalized ladder operators, satisfying $[B(t), B^\dagger(t)] = 1$, as follows:

$$\rho'(t) = \exp\left\{\left[-\frac{\ln R^2}{2} - \frac{1}{2}\ln\left(1 + R^{-2}\left(\frac{1}{\hbar}\sqrt{XY - \frac{1}{4}\dot{X}^2} + \frac{1}{2}(1 - R^2)\right)\right)\right]\{B,\cdot,B\}\right\}$$
$$\times \exp\left[-\frac{1}{2}\ln\left(1 + R^{-2}\left(\frac{1}{\hbar}\sqrt{XY - \frac{1}{4}\dot{X}^2} + \frac{1}{2}(1 - R^2)\right)\right)\{B^\dagger,\cdot,B^\dagger\}\right]\rho(0) \tag{45}$$

where

$$B = \sqrt{\frac{a}{2\hbar(ab - \text{Re}^2(c))^{1/2}}} e^{i\theta} q + \sqrt{\frac{b}{2\hbar(ab - \text{Re}^2(c))^{1/2}}} e^{i\phi} p \qquad (46)$$

with

$$a = \frac{-m \ln R^2}{2\hbar^2 R^2 (1 - R^2)} \left( \ddot{A}^2 X - \dot{A}\ddot{A}\dot{X} + \dot{A}^2 Y \right) \qquad (47)$$

$$b = \frac{-\ln R^2}{2m\hbar^2 R^2 (1 - R^2)} \left( A^2 Y - A\dot{A}\dot{X} + \dot{A}^2 X \right) \qquad (48)$$

$$c = \frac{\ln R^2}{4\hbar^2 R^2 (1 - R^2)} \left( 2A\dot{A}Y - A\ddot{A}\dot{X} + 2\dot{A}\ddot{A}X - \dot{A}^2 \dot{X} - i\hbar R^2 (1 - R^2) \right) \qquad (49)$$

and

$$\phi - \theta = \tan^{-1}\left[ \frac{(ab - \text{Re}^2(c))^{1/2}}{-\text{Re}(c)} \right] \qquad (50)$$

such that $0 \le \phi - \theta \le \pi$, one angle variable remaining arbitrary [6]. We will come back to Eq. (45) in the Discussion section. Comparing to solution (34), the conditions $w_1 \ge 0$, $w_2 \ge 0$ and $w_1 w_2 - w_3^2 \ge 0$ correspond respectively to $Y \ge 0$, $X \ge 0$ and $XY - \frac{\dot{X}^2}{4} \ge 0$, which are all true (the latter by the Cauchy-Schwarz inequality). However, consider the condition $w_4 > 0$, which for Ullersma's model corresponds to $-\ln R^2 > 0$ or, equivalently, $1 - R^2 > 0$, and which represents positive dissipation. One might intuit that for small $N$ or perhaps a peculiar density of states, the reservoir might give rise to negative dissipation, and indeed for Ullersma's model, one can show that $1 - R^2 < 0$ is possible at some times in the absence of additional restrictions. Consequently, one cannot apply the theorem mentioned in this section to derive the uncertainty relation (28) for Ullersma's model without such additional restrictions. Another derivation of the uncertainty relation (28) provided in Appendix 1 that, like the work in [2,3], stems from the conventional uncertainty principle, must also exclude negative dissipation.

One can understand physically why the condition $w_4 < 0$ must be excluded in such analyses. Under assumptions that include $w_4(t') > 0$ at some time $t' > 0$, it can be seen from the proof of the aforementioned theorem that if the exponent of Eq. (34) cannot be placed in Lindblad form, then the uncertainty principle $\langle q^2 \rangle \langle p^2 \rangle - \langle qp + pq \rangle^2 / 4 \ge \hbar^2 / 4$ is violated for some initial condition of the system of interest [12]. On the other hand, it is well understood that the uncertainty principle is maintained by the fluctuations of the reservoir that tend to counteract positive dissipation that reduces uncertainty [17]. With these remarks in mind, instead of positive dissipation with $w_4 > 0$, consider negative dissipation described by the equation

$$\frac{d\rho}{dt} = i\tilde{w}(\{q, \rho, p\} - \{p, \rho, q\}) \tag{51}$$

where the factor $\tilde{w}$ is less than zero and, for simplicity, is time independent. Because such an equation feeds energy into the system, one might guess that fluctuations are not needed to preserve the uncertainty principle, and this guess would be borne out: in addition to preserving norm and Hermiticity, Eq. (51) also preserves the uncertainty principle, yet it is not of Lindblad form [18]. The upshot is that if $w_4(t') < 0$, then the uncertainty principle can be maintained even if relation (39) does not hold.

Before ending this section, we make some remarks about the uncertainty relation (43) and the conventional uncertainty principle in two forms, assuming the class of master equations (33) and that relations (37) and (38) hold. As shown in Appendix 2, it is possible for a master equation to contradict relation (43) for some time intervals even though $\langle q^2\rangle\langle p^2\rangle \geq \hbar^2/4$ is obeyed for all $t \geq 0$ and all allowable initial conditions. On the other hand, a necessary and sufficient condition for $\langle q^2\rangle\langle p^2\rangle - \langle qp + pq\rangle^2/4 \geq \hbar^2/4$ to hold for all $t \geq 0$ and all allowable initial conditions is that relation (43) hold for all $t \geq 0$, necessity following from the results in Section 2 and Ref. [12]. Finally, if $e^{-w_4(t)}Trq^2\tilde{\mathcal{U}}_{rev}(t)\rho(0)$ and similar expressions for the other two reversible moments approach zero in the long time limit, then the uncertainty relation (43) and the latter form of the uncertainty principle become manifestly the same in this limit.

5. *Discussion*

We now interpret and discuss the significance of some of the foregoing results. With reference to Ullersma's model, assuming that the initial total state is factorized with the reservoir in thermal equilibrium and that $R^2 > 0$, the evolution of $\rho(t)$ is given by Eq. (44). If we pretend that we could just turn off the diffusion (i.e., just set $X, \dot{X}$ and $Y$ to zero in Eq. (44)), we would get $\langle q_0^2(t)\rangle = R^2(t)Tr(q_0^2\tilde{U}(t)\rho\tilde{U}^\dagger(t))$. Thus, $(\delta q)^2$ is the second moment stripped of the contribution from dissipation and reversible evolution, or what is the same, the part of the second moment that arises just from diffusion. Similar statements can be made about $(\delta p)^2$ and $C_\delta$. Moreover, since the right-hand side of the inequality (28) is related to dissipation, we confirm that inequality (28) is a type of dynamic fluctuation-dissipation relation.

If the total initial state is uncorrelated with the reservoir in thermal equilibrium, the quantity

$$D(t) \equiv (\delta q)^2(\delta p)^2 - \frac{1}{4}C_\delta^2 - \frac{\hbar^2}{4}(1 - R^2)^2 \tag{52}$$

only depends on $kT$ and the parameters of the total Hamiltonian and not on the initial condition of the system of interest. Moreover, the value of $D(t)$ provides information about the magnitude of the coefficients of $\{B, \cdot, B\}$ and $\{B^\dagger, \cdot, B^\dagger\}$ in Eq. (45). This last remark is related to the known result that $-r_1\{q, \cdot, q\} + r_3\{q, \cdot, p\} + r_3^*\{p, \cdot, q\} - r_2\{p, \cdot, p\}$, with $r_1 r_2 = |r_3|^2$, can be written in Lindblad form with just one ladder operator [20]. The initial value of $D$ is zero and evolves to some quantity in the range $[0,\infty)$. Suppose that at time $t'$ we were to again have $D = 0$. Then, it can be easily checked that the first exponent in Eq. (45) vanishes at $t'$ if $1 - R^2 > 0$. In such case, there exists an initial, pure system state (in the language of quantum optics, a two-photon coherent state [21]) that evolves to another pure system state (another two-photon coherent state) at time $t'$ [6]. This is related to Hasse's [22] condition for pure state evolution for autonomous master equations. In this sense, $D(t)$ is a measure of the ability of $H_T$ to maintain coherence in the reduced dynamics at time $t$, $D = 0$ indicating that completely coherent evolution is possible for some initial states. A kind of inverse also exists: conditions can be formulated that ensure that $D(t) = 0$ [23]. A corresponding result for autonomous systems was previously obtained in Ref. [5] (see Eq. (12) ff. therein). Due to Eq. (16) we may also note that if, at some time, $D(t)$ is greater than zero, it cannot be zero at this time at some higher temperature.

The foregoing statements might not apply when $D(t)$ is subjected to popular approximations. For example, if the continuum limit is adopted using Ullersma's spectral strength [1] with $\alpha \geq 3\Gamma$, and if the zero point contribution to the reservoir energy is neglected, then $D(t)$ necessarily falls below zero at some time, even if all other temperature dependent terms of the energy are maintained [12]. Thus, for this case, invoking only the continuum limit and the high temperature approximation $\hbar\omega\left(\left(e^{\hbar\omega/kT}-1\right)^{-1}+1/2\right) \to kT$ leads to negative $D(t)$ for at least some times and initial states.

The specific FD inequality (17) may be derived by examining the reduced density operator or expectation values when the initial state is given by the factored form (1) (cf. Section 4 and Appendix 1, respectively). To obtain other FD inequalities by a similar procedure may be difficult for total initial states that are correlated because in such case expectation values and especially the reduced density operator may not be readily available. Fortunately, the class of FD inequalities that are encompassed by relation (15) was derived without reference to any particular total initial state and therefore affords an opportunity to investigate the time evolution of fluctuations and dissipation in circumstances that may be intractable using traditional methods.

*Appendix 1*

Here we show that another derivation of (28), based on the uncertainty principle, requires us to exclude negative dissipation (i.e., $1 - R^2 < 0$), as in Section 4. The uncertainty principle has previously been used to derive inequalities involving the second moments, $R$, and environment-induced diffusion coefficients by considering a factorized total initial state with the reservoir in a Gaussian state [2,3]. One particular result of this last reference (Eq. (40) therein), specialized to the zeroth oscillator (with canonical variables $q_0$, $p_0$) of the system of interest, furnishes

$$\begin{pmatrix} \langle q_0^2 \rangle(t) & \langle q_0 p_0 + p_0 q_0 \rangle(t)/2 \\ \langle q_0 p_0 + p_0 q_0 \rangle(t)/2 & \langle p_0^2 \rangle(t) \end{pmatrix} \geq -\frac{i\hbar}{2}\tilde{R}(t) J \tilde{R}^T(t) + S(t) \qquad (53)$$

where

$$\tilde{R}(t) = \begin{pmatrix} A & A/m \\ m\ddot{A} & \dot{A} \end{pmatrix}, \quad J = \begin{pmatrix} 0 & 1 \\ -1 & 0 \end{pmatrix} \text{ and } S(t) = \begin{pmatrix} X/m & \dot{X}/2 \\ \dot{X}/2 & mY \end{pmatrix}.$$

If, as claimed therein, equality in expression (53) can be achieved for some pure states, then the uncertainty relation (28) could be derived by writing $-i\hbar \tilde{R}(t) J \tilde{R}^T(t)/2 + S(t) + i\hbar J/2 \geq 0$, considering the determinant of the left-hand side and then using Eqs. (20)-(25). Unfortunately, equality in expression (53) implies unphysical initial moments. Thus, we take the following somewhat different tack more akin to the approach leading to expression (4.10) of Ref. [2].

Eqs. (20)-(25) yield

$$\langle q_0^2 \rangle(t)\langle p_0^2 \rangle(t) - \frac{1}{4}\langle q_0 p_0 + p_0 q_0 \rangle^2(t) =$$

$$XY - \frac{1}{4}\dot{X}^2 + R^4\left( Tr q_0^2 \tilde{U}(t)\rho \tilde{U}^\dagger(t) Tr p_0^2 \tilde{U}(t)\rho \tilde{U}^\dagger(t) - \frac{1}{4}\left( Tr(q_0 p_0 + p_0 q_0)\tilde{U}(t)\rho \tilde{U}^\dagger(t) \right)^2 \right)$$

$$+ R^2\left( m_0 Y Tr q_0^2 \tilde{U}(t)\rho \tilde{U}^\dagger(t) - \frac{1}{2}\dot{X} Tr(q_0 p_0 + p_0 q_0)\tilde{U}(t)\rho \tilde{U}^\dagger(t) + \frac{X}{m_0} Tr p_0^2 \tilde{U}(t)\rho \tilde{U}^\dagger(t) \right)$$

$$\geq \frac{\hbar^2}{4},$$
(54)

the last inequality being valid due to the uncertainty principle, which holds if $\rho(t)$ is positive (a condition which may not be easy to check if approximations are introduced). Let's assume $4XY - \dot{X}^2 > 0$ and for the three quantities $Tr q_0^2 \tilde{U}(t)\rho \tilde{U}^\dagger(t)$, $\frac{1}{2}Tr(q_0 p_0 + p_0 q_0)\tilde{U}(t)\rho \tilde{U}^\dagger(t)$ and 

$Tr p_0^2 \tilde{U}(t)\rho \tilde{U}^\dagger(t)$ choose $\dfrac{\hbar X}{m_0\sqrt{4XY - \dot{X}^2}}, \dfrac{\hbar \dot{X}}{2\sqrt{4XY - \dot{X}^2}}$ and $\dfrac{\hbar m_0 Y}{\sqrt{4XY - \dot{X}^2}}$. If we assume $R^2 > 0$,

we obtain

$$\sqrt{XY - \dot{X}^2/4} \geq \hbar(1 - R^2)/2.$$
(55)

Provided $1 - R^2 \geq 0$, we arrive at Eq. (17). Finally, heeding Eqs. (20)-(25) leads to the uncertainty relation (28), although, as previously noted, this relation cannot be reached with the foregoing proof if we assume $1 - R^2 < 0$.

*Appendix 2*

For $t \geq 0$, suppose $\rho(t)$ is given by Eq. (34) where $w_1(t) \geq 0$, $w_2(t) \geq 0$, $w_3(t)$ and $w_4(t)$ are real continuous functions that vanish at $t = 0$, such that $0 \leq w_1(t)w_2(t) - w_3^2(t)$. Suppose further that all three functions $w_1(t')$, $w_2(t')$ and $w_4(t')$ are positive for $t'>0$. Functions satisfying these properties and the additional inequality $w_1(t')w_2(t') - w_3^2(t') < \left(\dfrac{e^{w_4(t')}-1}{4\hbar}\right)^2$ (this is the negation of Eq. (39)) for $t'>0$ can be found such that $\langle q^2 \rangle(t)\langle p^2 \rangle(t) \geq \hbar^2/4$ holds. Thus, the uncertainty relation (43) is not a consequence of the uncertainty principle $\langle q^2 \rangle(t)\langle p^2 \rangle(t) \geq \hbar^2/4$. We prove this by choosing $w_1(t)w_2(t) = w_3^2(t)$, as this ensures $w_1(t')w_2(t') - w_3^2(t') < \left(\dfrac{e^{w_4(t')}-1}{4\hbar}\right)^2$ for $t'>0$ under the hypotheses. We will show that by further choosing

$$w_1(t')w_2(t') = \left(\frac{1}{4\hbar}\right)^2 \left(e^{2w_4(t')} - 1\right), \tag{56}$$

the uncertainty principle $\langle q^2 \rangle(t)\langle p^2 \rangle(t) \geq \hbar^2/4$ holds, which will complete the proof. To this end, let us use Eqs. (40) and (41), to re-write
$\langle q^2 \rangle(t)\langle p^2 \rangle(t) - \hbar^2/4 \geq 0$
$\Leftrightarrow$
$e^{-w_4(t)}\left[Trq^2 \mathcal{U}_{rev}(t)\rho(0) + 2\hbar^2 w_2(t)\right] e^{-w_4(t)}\left[Trp^2 \mathcal{U}_{rev}(t)\rho(0) + 2\hbar^2 w_1(t)\right] - \hbar^2/4 \geq 0$
$\Leftrightarrow$

$$w_1(t)(Trq^2\mathcal{U}_{rev}(t)\rho(0))^2 + \left[2\hbar^2\left(w_1(t)w_2(t) - \left(\frac{e^{w_4(t)}-1}{4\hbar}\right)^2\right) - \frac{(e^{w_4(t)}-1)}{4}\right]Trq^2\mathcal{U}_{rev}(t)\rho(0) + \frac{\hbar^2}{4}w_2(t)$$

$$+ \left(w_2(t) + \frac{Trq^2\mathcal{U}_{rev}(t)\rho(0)}{2\hbar^2}\right)\left(Trq^2\mathcal{U}_{rev}(t)\rho(0)Trp^2\mathcal{U}_{rev}(t)\rho(0) - \frac{\hbar^2}{4}\right) \geq 0$$

The quantity in the square brackets vanishes on account of Eq. (56), and the remaining three terms on the left-hand side of the last inequality are non-negative, which completes the proof.

pertain to Ullersma's density of states, one can show that under certain conditions that maintain the positivity of $\rho(t)$, $XY - \dot{X}^2/4 - \hbar^2(1-R^4)/4$ always drops below zero at short times.